\documentclass[epjCONF]{svjour}
\usepackage{graphics}
\usepackage{epsfig}
\usepackage[varg]{txfonts} 
\usepackage[latin1]{inputenc}
\session-title{HCBM 2010-- Hot $\&$ Cold Baryonic Matter 2010}
\begin{document}
\title{Interplay between chiral and deconfinement phase transitions}
\author{Fukun Xu\inst{1} \and
Tamal K. Mukherjee\inst{1,2} \and Huan Chen\inst{3} \and Mei
Huang\inst{1,2} }
\institute{Institute of High Energy Physics, Chinese Academy of
Sciences, Beijing, China  \and Theoretical Physics Center for
Science Facilities, Chinese Academy of Sciences, Beijing, China \and
Istituto Nazionale di Fisica Nucleare (INFN), Sezione di Catania,
Italy}
\abstract{By using the dressed Polyakov loop or dual chiral
condensate as an equivalent order parameter of the deconfinement
phase transition, we investigate the relation between the chiral and
deconfinement phase transitions at finite temperature and density in
the framework of three-flavor Nambu--Jona-Lasinio (NJL) model. It is
found that in the chiral limit, the critical temperature for chiral
phase transition coincides with that of the dressed Polyakov loop in
the whole $(T,\mu)$ plane. In the case of explicit chiral symmetry
breaking, it is found that the phase transitions are flavor
dependent. For each flavor, the transition temperature for chiral
restoration $T_c^{\chi}$ is smaller than that of the dressed
Polyakov loop $T_c^{{\cal D}}$ in the low baryon density region
where the transition is a crossover, and, the two critical
temperatures coincide in the high baryon density region where the
phase transition is of first order. Therefore, there are two
critical end points, i.e, $T_{CEP}^{u,d}$ and $T_{CEP}^{s}$ at
finite density. We also explain the feature of $T_c^{\chi}=T_c^{\cal
D}$ in the case of 1st and 2nd order phase transitions, and
$T_c^{\chi}<T_c^{\cal D}$ in the case of crossover, and expect this
feature is general and can be extended to full QCD theory.
} 
\maketitle
\section{Introduction}
\label{intro}

The interplay between chiral symmetry breaking and confinement as
well as the chiral and deconfinement phase transitions at finite
temperature and density are of continuous interests
\cite{Polyakov:1978vu,'tHooft:1977hy,Casher:1979vw,Banks:1979yr,Hatta:2003ga,Mocsy:2003qw,Marhauser:2008fz,Braun:2007bx,Braun:2009gm}.
The two transitions are characterized by the breaking and
restoration of chiral and center symmetry, which are well defined in
two extreme quark mass limits, respectively. In the chiral limit
when the current quark mass is zero $m=0$, the chiral condensate is
the order parameter for the chiral phase transition. When the
current quark mass goes to infinity $m\rightarrow \infty$, QCD
becomes pure gauge $SU(3)$ theory, which is center symmetric in the
vacuum, and the usually used order parameter is the Polyakov loop
\cite{Polyakov:1978vu}.

The relation between the chiral and deconfinement phase transitions
has attracted more interest recently in studying the phase diagram
at high baryon density region \cite{QCD-phase}. It is conjectured in
Ref. \cite{McLerran:2007qj} that in large $N_c$ limit, a confined
but chiral symmetric phase, which is called quarkyonic phase can
exist in the high baryon density region. It is very interesting to
study whether this quarkyonic phase can survive in real QCD phase
diagram, and how it competes with nuclear matter and the color
superconducting phase \cite{CSC}. (However, it is worthy of noticing
that in Ref. \cite{Panero:2009tv}, it is found that at zero chemical
potential, the lattice results for the thermodynamical properties
have a very mild dependence on the number of colors.)

Lattice QCD at the current stage cannot go to very high baryon
density. For zero chemical potential, previous lattice results show
that the chiral and deconfinement phase transitions occur at the
same temperature, e.g, in Ref.
\cite{Kogut:1982rt,Fukugita:1986rr,Karsch:1994hm,Digal:2000ar,Digal:2002wn},
and also in review papers \cite{Karsch:2001cy,Laermann:2003cv}. In
recent years, three lattice groups, MILC group
\cite{Bernard:2004je}, RBC-Bielefeld group \cite{Cheng:2006qk} and
Wuppertal-Budapest group \cite{Aoki:2006br,Aoki:2009sc,Fodor:2009ax}
have studied the chiral and deconfinement phase transition
temperatures with almost physical quark masses. The RBC-Bielefeld
group claimed that the two critical temperatures for $N_f=2+1$
coincide at $T_c=192(7)(4) {\rm MeV}$. The Wuppetal-Budapest group
found that for the case of $N_f=2+1$, there are three transition
temperatures, the transition temperature for chiral restoration of
$u,d$ quarks $T_c^{\chi(ud)}= 151(3)(3)~{\rm MeV}$, the transition
temperature for chiral restoration of $s$ quark
$T_c^{\chi(s)}=175(2)(4){\rm MeV}$ and the deconfinement transition
temperature $T_c^d=176(3)(4) {\rm MeV}$. From this result, we can
read that the critical temperatures for different transitions are
different. According to the Wuppetal-Budapest group, this is the
consequence of the crossover nature.

In the framework of QCD effective models, there is still no
dynamical model which can describe the chiral symmetry breaking and
confinement simultaneously. The main difficulty of effective QCD
model to include confinement mechanism lies in that it is difficult
to calculate the Polyakov loop analytically. Currently, the popular
models used to investigate the chiral and deconfinement phase
transitions are the Polyakov Nambu-Jona-Lasinio model (PNJL)
\cite{Fukushima:2003fw,Ratti:2005jh,Sasaki:2006ww,Ghosh:2006qh,Fu:2007xc,Zhang:2006gu,Fukushima:2008wg,Abuki:2008nm}
and Polyakov linear sigma model (PLSM)
\cite{Schaefer:2007pw,Mao:2009aq}. However, the shortcoming of these
models is that the temperature dependence of the Polyakov-loop
potential is put in by hand from lattice result, which cannot be
self-consistently extended to finite baryon density. Recently,
efforts have been made in Ref.\cite{Kondo:2010ts} to derive a
low-energy effective theory for confinement-deconfinement and
chiral-symmetry breaking/restoration from first-principle.

Recent investigation revealed that quark propagator, heat kernels
can also act as an order parameter as they transform non trivially
under the center transformation related to deconfinement transition
\cite{Synatschke:2007bz,Synatschke:2008yt,Bilgici:2008ui}. But the
exciting result is the behavior of spectral sum of the Dirac
operator under center transformation
\cite{Synatschke:2008yt,Gattringer:2006ci,Bruckmann:2006kx,Bilgici:2008qy}.
A new order parameter, called dressed Polyakov loop has been defined
which can be represented as a spectral sum of the Dirac operator
\cite{Bilgici:2008qy}. It has been found the infrared part of the
spectrum particularly plays a leading role in confinement
\cite{Synatschke:2008yt}. This result is encouraging since it gives
a hope to relate the chiral phase transition with the
confinement-deconfinement phase transition. The order parameter for
chiral phase transition is related to the spectral density of the
Dirac operator through Banks-Casher relation \cite{Banks:1979yr}.
Therefore, both the dressed Polyakov loop and the chiral condensate
are related to the spectral sum of the Dirac operator.

Behavior of the dressed Polyakov loop is mainly studied in the
framework of Lattice gauge theory
\cite{Bruckmann:2008br,Bilgici-thesis}. Apart from that, studies
based on Dyson-Schwinger equations
\cite{Fischer:2009wc,Fischer:2009gk,Fischer:2010fx} and PNJL model
\cite{Kashiwa:2009ki} have been carried out. In those studies the
role of dressed Polyakov loop as an order parameter is discussed at
zero chemical potential. In this paper, we show our results
\cite{ourwork} of investigating the QCD phase diagram at finite
temperature and density by using the dressed Polyakov loop as an
equivalent order parameter in the Nambu-Jona-Lasinio (NJL) model .
It is known that he NJL model lacks of confinement and the gluon
dynamics is encoded in a static coupling constant for four point
contact interaction. However, assuming that we can read the
information of confinement from the dual chiral condensate, it would
be interesting to see the behavior of the dressed Polyakov loop in a
scenario without any explicit mechanism for confinement.

In this paper, we show the phase transitions in the two-flavor and
three-flavor NJL models in $(T,\mu)$ plane in chiral limit as well
for small quark mass limit. This paper is organized as follows: We
introduce the dressed Polyakov loop as an equivalent order parameter
of confinement deconfinement phase transition and the NJL model in
Sec. \ref{DPL-section}. Then in Sec.\ref{NM-section}, we show the
results of  two-flavor QCD phase diagram in $T-\mu$ plane in the
chiral limit and in the case of explicit chiral symmetry breaking,
respectively. We offer an analysis on the relation between the
chiral and deconfinement phase transitions in Sec.
\ref{two-tc-section}. In Sec.\ref{SU3-section}, we show the phase
diagram at finite temperature and density for three-flavor case. At
the end, we give the conclusion and discussion.

\section{Dressed Polyakov loop and the NJL model}
\label{DPL-section}

We firstly introduce the dressed Polyakov loop. To do this we have
to consider a $U(1)$ valued boundary condition for the fermionic
fields in the temporal direction instead of the canonical choice of
anti-periodic boundary condition,
\begin{equation}
\psi(x,\beta) = e^{-i \phi} \psi(x,0),
\end{equation}
where $0\leq \phi < 2\pi $ is the phase angle and $\beta$ is the
inverse temperature.

Dual quark condensate $\Sigma_n$ is then defined by the Fourier
transform (w.r.t the phase $\phi$) of the general boundary condition
dependent quark condensate
\cite{Bilgici:2008qy,Bruckmann:2008br,Bilgici-thesis},
\begin{equation}
\Sigma_n =-{\int_0}^{2\pi} \frac{d\phi}{2\pi} e^{-i n \phi}
\langle\bar{\psi} \psi\rangle_\phi, \label{eq.dpl}
\end{equation}
where $n$ is the winding number.

Particular case of $n=1$ is called the dressed Polyakov loop which
transforms in the same way as the conventional thin Polyakov loop
under the center symmetry and hence is an order parameter for the
deconfinement transition
\cite{Bilgici:2008qy,Bruckmann:2008br,Bilgici-thesis}. It reduces to
the thin Polyakov loop and to the dual of the conventional chiral
condensate in infinite and zero quark mass limits respectively,
i.e., in the chiral limit $m \rightarrow 0$ we get the dual of the
conventional chiral condensate and in the $m\rightarrow \infty$
limit we have thin Polyakov loop
\cite{Bilgici:2008qy,Bruckmann:2008br,Bilgici-thesis}.

The Lagrangian of three-flavor NJL model \cite{NJL-report} is given
as
\begin{eqnarray}
  {\cal{L}} & = & \bar{\psi}(i\gamma^{\mu}\partial_{\mu}-m)\psi
    + G_s \sum_a \Big\{ (\bar{\psi}\tau_a\psi)^2 + (\bar{\psi}i\gamma_5\tau_a\psi)^2
    \Big\} \nonumber \\
   & - & K \Big\{ {\rm Det}_f[\bar{\psi}(1+\gamma_5)\psi] + {\rm Det}_f[\bar{\psi}(1-\gamma_5)\psi]
   \Big\}.
\label{Lagr-flavor-3}
\end{eqnarray}
Where $\psi=(u,d,s)^T$ denotes the transpose of the quark field, and
$m={\rm Diag}(m_u,m_d,m_s)$ is the corresponding mass matrix in the
flavor space. $\tau_a$ with $a=1,\cdots,N_f^2-1$ are the eight
Gell-Mann matrices, and ${\rm Det}_f$ means determinant in flavor
space. The last term is the standard form of the 't Hooft
interaction, which is invariant under $SU(3)_L\times SU(3)_R\times
U(1)_B$ symmetry, but breaks down the $U_A(1)$ symmetry.

The $\phi$ dependent thermodynamic potential in the mean field level
is given as following:
\begin{eqnarray}
\Omega_{\phi}  =  \sum_f{\Omega_{\phi,M_f}} +
2G_s\sum_f{\langle\sigma\rangle_{\phi,f}^2}
    -  4K\langle\sigma\rangle_{\phi,u} \langle\sigma\rangle_{\phi,d}
   \langle\sigma\rangle_{\phi,s},
\label{thermal-pot-flavor-3} \end{eqnarray} with
\begin{eqnarray}
\Omega_{\phi,M_f} & = & -2N_c {\int_\Lambda} \frac{d^3 p}{(2\pi)^3}
    \Big[E_{p,f}+{1\over\beta}ln(1+e^{-\beta E_{p,f}^-}) \nonumber \\
    & + & {1\over\beta}ln(1+e^{-\beta
    E_{p,f}^+})\Big].
\end{eqnarray}
Where the sum is in the flavor space,
$E_{p,f}=\sqrt{p^2+M^2_{\phi,f}}$ and $E_{p,f}^{\pm}=E_{p,f} \pm
[\mu+i(\phi-\pi)T]$, with the constituent quark mass
\begin{equation}
M_{\phi,i}=m_i-4G_s\langle\sigma\rangle_{\phi,i}+2K\langle\sigma\rangle_{\phi,j}\langle\sigma\rangle_{\phi,k},
\end{equation}
where $(i,j,k)$ is the quark flavor indices $(u,d,s)$, and
$\langle\sigma\rangle_{\phi,f}=\langle\bar{\psi}_f\psi_f\rangle_{\phi}$.

\section{Phase diagram for two-flavor case}
\label{NM-section}

We firstly show the results in the two-flavor case and consider the
isospin symmetric limit, i.e, we take $N_f=2$, $K=0$ and $m_u=m_d$
in Eq.(\ref{Lagr-flavor-3}). The thermodynamic potential contains
imaginary part. We take only the real part of the potential and the
imaginary phase factor is not considered in this work. The mean
field $\langle\sigma\rangle_{\phi}$ is obtained by minimizing the
potential for each value of $\phi \in [0,2\pi)$ for fixed T and
$\mu$. The conventional chiral condensate is
$\langle\sigma\rangle_{\pi}=\langle\bar{\psi}\psi\rangle_{\pi}$. The
dressed Polyakov loop $\Sigma_{1}$ is obtained by integrating over
the angle. The values of the parameters $\Lambda$ and $G_s$ are
taken as $0.6315 {\rm GeV}$ and $5.498 {\rm GeV}^{-2}$,
respectively.

\begin{figure}[thbp]
\epsfxsize=6.5 cm \epsfysize=6 cm \epsfbox{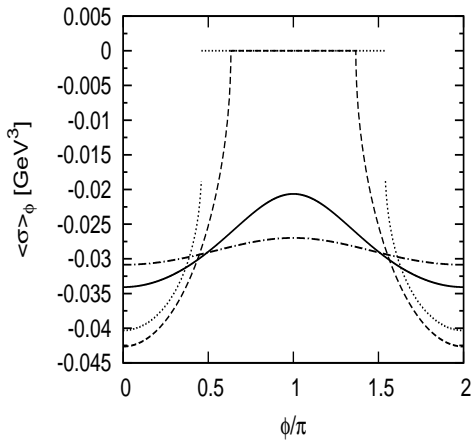} \caption{Angle
variation of $\langle\sigma\rangle_\phi$ for different values of
temperatures and chemical potentials in the case of chiral limit.
The solid line corresponds to $T= 150 {\rm MeV}, \mu = 100 {\rm
MeV}$, dashed line corresponds to $T= 250 {\rm MeV}, \mu = 100 {\rm
MeV}$, dash-dotted line corresponds to $T= 40 {\rm MeV}, \mu = 300
{\rm MeV}$, and dot corresponds to $T= 150 {\rm MeV},  \mu = 300
{\rm MeV}$. } \label{fig-angle-m0}
\end{figure}

\begin{figure}[thbp]
\epsfxsize=6.5 cm \epsfysize=6 cm \epsfbox{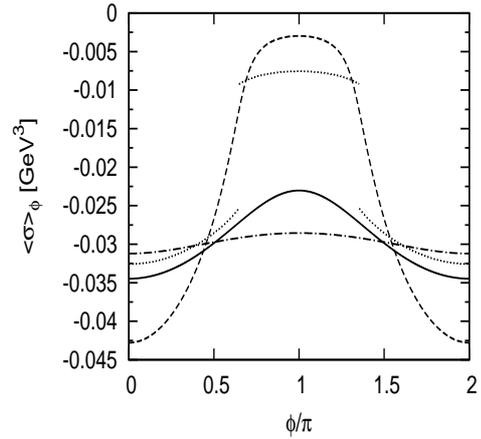} \caption{Angle
variation of $\langle\sigma\rangle_\phi$ for different values of
temperatures and chemical potentials in the case of $m=5.5 {\rm
MeV}$. The solid line corresponds to $T= 150 {\rm MeV} , \mu = 100
{\rm MeV}$, dashed line corresponds to $T= 250 {\rm MeV}, \mu = 100
{\rm MeV}$, dash-dotted line corresponds to $T= 20 {\rm MeV}, \mu =
340 {\rm MeV}$ and dotted line corresponds to $T= 40 {\rm MeV}, \mu
= 340 {\rm MeV}$. } \label{fig-angle-m055}
\end{figure}

We investigate phase transitions for two cases, i.e., in the chiral
limit and with explicit chiral symmetry breaking. Fig.
\ref{fig-angle-m0} and \ref{fig-angle-m055} show the behavior of the
angle dependence of the general chiral condensate for various
chemical potentials and temperatures for $m = 0$ and $m = 5.5 {\rm
MeV}$, respectively. The four curves presented in each figure
represent two temperatures above and below the critical temperature
for two particular values of the chemical potential. Same
qualitative features have been found for both the quark masses. The
variation is symmetrical around $\phi = \pi$ as reported in other
studies \cite{Fischer:2009wc,Kashiwa:2009ki}.

Almost no variation with respect to angle is found for low
temperatures. As the temperature increases the variation over the
angle grows. We expect the absolute value of the chiral condensate
decreases with the increase of temperature. However, from the
figure, this conventional behavior of the chiral condensate with
temperature only persists up to a certain angle, beyond which the
opposite behavior is observed. The plateau around $\phi = \pi$ is
more flat above $T_c$ in case of zero current quark mass and its
value is consistent with the expectation of complete restoration of
chiral symmetry in the chiral limit.

\begin{figure}[thbp]
\epsfxsize=7.5 cm \epsfysize=6.5cm \epsfbox{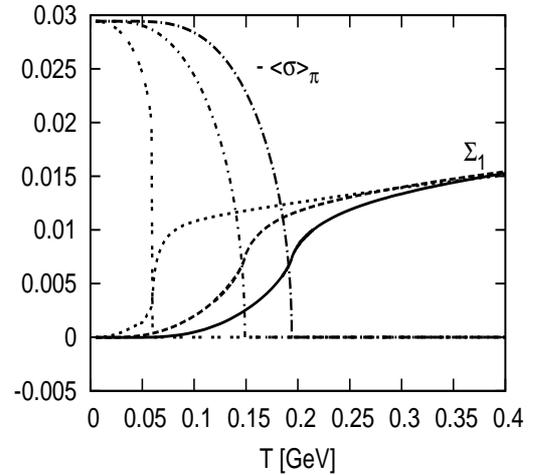}
 \caption{The conventional chiral condensate $-\langle\sigma\rangle_{\pi}$ and
the dressed Polyakov loop $\Sigma_1$ as functions of temperature for
different values of the chemical potentials in the chiral limit.
Here, $-\langle\sigma\rangle_{\pi}$ and $\Sigma_1$ both are measured
in $[GeV^3]$. From right to left the values of the chemical
potential are $0, 200, 300 {\rm MeV}$, respectively. }
 \label{fig-dplm0}
\end{figure}

\begin{figure}[thbp]
\epsfxsize=7.5 cm \epsfysize=6.5cm \epsfbox{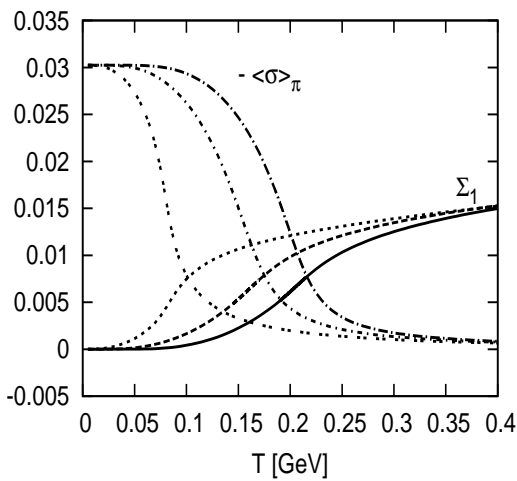}
 \caption{The conventional chiral condensate $-\langle\sigma\rangle_{\pi}$ and
the dressed Polyakov loop $\Sigma_1$ as functions of temperature for
different values of the chemical potentials in the case of explicit
chiral symmetry breaking $m=5.5 {\rm MeV}$ and
$-\langle\sigma\rangle_{\pi}$, $\Sigma_1$ are measured in $[GeV^3]$.
From right to left the values of the chemical potential are $0, 200,
300 {\rm MeV}$, respectively.}
 \label{fig-dplm055}
\end{figure}

Fig. \ref{fig-dplm0} and \ref{fig-dplm055} show the behavior of the
conventional chiral condensate $-\langle\sigma\rangle_{\pi}$ and the
dressed Polyakov loop $\Sigma_1$ at different chemical potentials as
functions of temperature for $m=0$ and $m=5.5 {\rm MeV}$,
respectively. For both cases, it is observed there are three
temperature regions for $-\langle\sigma\rangle_{\pi}$ and
$\Sigma_1$. For $-\langle\sigma\rangle_{\pi}$, at smaller
temperatures it remains constant at a value corresponding to the
value of the conventional chiral condensate in the vacuum, then it
rapidly decreases in a small window of temperature and eventually
almost saturates to a lower value. The rapid decreasing occurs at
different temperatures for different values of the chemical
potentials. On the other hand the behavior for the dressed Polyakov
loop is just the opposite. It remains almost zero for small
temperatures and then rises rapidly, finally saturates to a high
value which varies very slowly with temperatures. The almost zero
value of $\Sigma_1$ for small temperatures is due to the fact that
the $U(1)$ boundary condition dependent general quark condensate
nearly does not vary with the angle $\phi$ for small temperatures
(see Eq.~\ref{eq.dpl}).

For finite quark mass, near the critical temperature region, both
$-\langle\sigma\rangle_{\pi}$ and $\Sigma_1$ change more slowly than
those in the case of chiral limit.

\begin{figure}[thbp]
\epsfxsize=7.5 cm \epsfysize=6.5cm \epsfbox{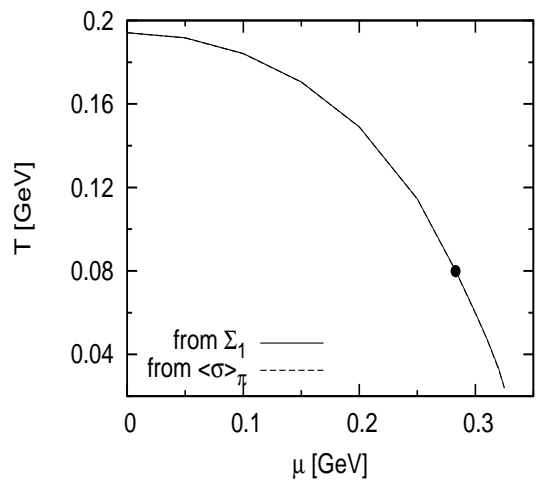}
\caption{Two-flavor phase diagram in the $T-\mu$ plane for the case
of chiral limit. The solid line is the critical line for $\Sigma_1$,
and the dashed line is the critical line for conventional chiral
phase transition. The solid circle indicates the critical point for
chiral phase transition. } \label{fig-phasediag-m0}
\end{figure}

\begin{figure}[thbp]
\epsfxsize=7.5 cm \epsfysize=6.5cm \epsfbox{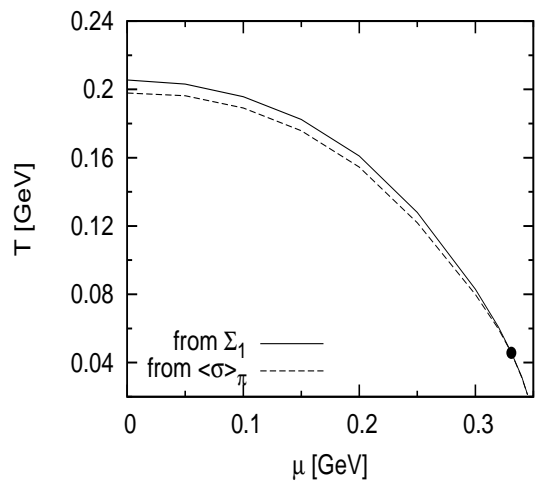}
\caption{Two-flavor phase diagram in the $T-\mu$ plane for the case
of $m=5.5 {\rm MeV}$. The solid line is the critical line for
$\Sigma_1$, and the dashed line is the critical line for
conventional chiral phase transition. The solid circle indicates the
critical end point for chiral phase transition.}
\label{fig-phasediag-m055}
\end{figure}

Fig. \ref{fig-phasediag-m0} and \ref{fig-phasediag-m055} show the
$T-\mu$ phase diagram for the case of $m=0$ and $m=5.5 {\rm MeV}$,
respectively. The transition temperatures are calculated from the
slope analysis of the conventional chiral condensate
$\langle\sigma\rangle_{\pi}$ and the dressed Polyakov loop. The
transition temperatures calculated from the conventional chiral
condensate represent the chiral phase transition temperature. On the
other hand the behavior of the dressed Polyakov loop is supposed to
indicate the deconfinement transition temperature. In our present
framework confinement is not accounted for. However, if we look at
the curves presented in figure~\ref{fig-dplm0} and
\ref{fig-dplm055}, they still show an order parameter like behavior.
We would like to point out here that for the same reason as stated
above we are not concerned here with the order of the phase
transtion from the dressed Polyakov loop and all the comments about
the order of the phase transition below are with respect to the
chiral phase transition only.

For $m = 0$ case, we find almost exact matching for the transition
temperatures calculated from these two quantities in the whole
$T-\mu$ plane as shown in Fig. \ref{fig-phasediag-m0}.

For the case of finite quark mass $m=5.5 {\rm MeV}$, it is observed
from Fig. \ref{fig-phasediag-m055} that the two critical
temperatures are different in the low baryon density region. The
difference however decreases from low to high chemical potential,
and the two critical temperatures start to match around the critical
end point for chiral phase transition. For zero chemical potential
and small current quark mass $m=5.5 {\rm MeV}$, we find about $7
{\rm MeV}$ difference between $T_c^{\chi}$ and $T_c^{\cal D}$, and
$T_c^{\chi}<T_c^{\cal D}$. Similar trend has been observed in
another study based on Dyson-Schwinger approach
\cite{Fischer:2009wc}, where they found chiral transition to occur
about $10-20 {\rm MeV}$ below the deconfinement transition. Though
these studies are not a complete one and these differences may be
due to the effects of crossover transition. As pointed out in
\cite{Fodor:2009ax} during crossover, different observables are
expected to behave differently and there is no way to define a
unique crossover temperature.

\begin{figure}[thbp]
\epsfxsize=7.5 cm \epsfysize=6.5cm \epsfbox{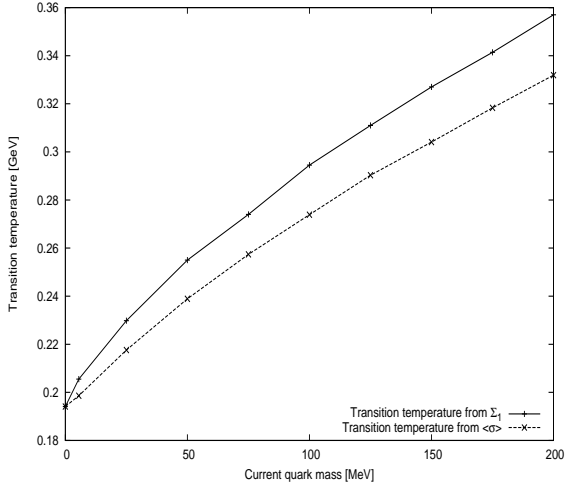} \caption{The
critical temperatures $T_c^{\chi}$ and $T_c^{\cal D}$ for different
current masses. } \label{fig-MassEffect}
\end{figure}

We extend our study further to see what happens if we increase the
current quark mass further. We find at zero chemical potential, the
difference between the transition temperatures calculated from
dressed Polyakov loop and conventional chiral condensate increases
as we increase the current quark mass (see
Fig.~\ref{fig-MassEffect}). Initially the difference is zero for
zero current quark mass but for $m=200 {\rm MeV}$ we find about $26
{\rm MeV}$ difference between the two temperatures. It is worthy of
mentioning that this result is just for illustrative purpose as
there are limitation of using NJL model with such a huge current
quark mass.

\section{The relation between $T_c^{\chi}$ and $T_c^{\cal D}$}
\label{two-tc-section}

In the following, we offer a possible understanding on the
simultaneity of the transition temperatures for 1st and 2nd order
chiral phase transitions and the apparent difference between the two
for the case of crossover.

\begin{figure}[thbp]
\epsfxsize=7.5 cm \epsfysize=6.5cm \epsfbox{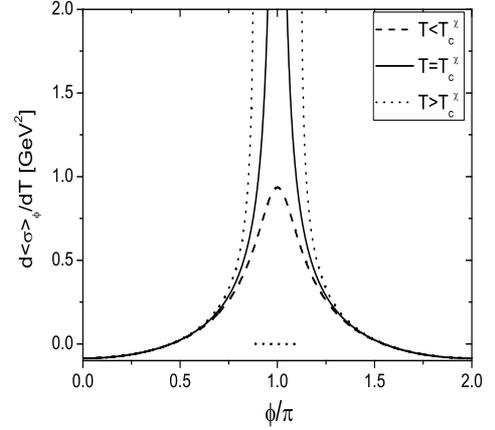}
\caption{Temperature derivative of the general chiral condensate
$\frac{d\langle\sigma\rangle_\phi}{dT}$ for $m=0$ and $\mu=0$ at
three temperature cases: $T < {T_c}^\chi$, $T = {T_c}^\chi$ and $T>
{T_c}^\chi$, where ${T_c}^\chi$ is the chiral transition
temperature. } \label{fig-firstderi-2order}
\end{figure}

As mentioned earlier the transition temperatures are determined from
the slope analysis of the conventional chiral condensate and the
dressed Polyakov loop. So let us look at the temperature derivative
of the general chiral condensate $d\langle\sigma\rangle_\phi/dT$ as
functions of $\phi$, the integral on which gives the temperature
derivative of the Dressed Polyakov loop
\begin{equation}
\frac{d\Sigma_1}{dT} =-{\int_0}^{2\pi} \frac{d\phi}{2\pi} e^{-i
\phi} \frac{d\langle\sigma\rangle_\phi}{dT}. \label{eq.dpl1}
\end{equation}
Fig. \ref{fig-firstderi-2order} shows
$d\langle\sigma\rangle_\phi/dT$ at different temperatures in the
cases of second order chiral phase transitions. Around $T_c^\chi$,
large values of $d\langle\sigma\rangle_\phi/dT$ appear around $\phi
=\pi$ and dominate the integral in Eq. (\ref{eq.dpl1}). Below
$T_c^\chi$ , $d\langle \sigma\rangle_\phi/dT$ around $\phi =\pi$
increases monotonously as temperature increases. As a result,
$d\Sigma_1/dT$ increases. Above $T_c^\chi$,
$d\langle\sigma\rangle_\phi/dT$ in the center region becomes zero
(or very small in the case with finite current quark mass) and the
region with large values of $d\langle \sigma\rangle_\phi/dT$
shrinks. Therefore, the integral in Eq. (\ref{eq.dpl1}) i.e.
$d\Sigma_1/dT$ decreases as temperature increases.  In all,
$d\Sigma_1/dT$ gets its maximum at $T_c^\chi$ and the two transition
temperatures $T_c^{\cal D}$ and $T_c^\chi$ coincide in the case of
second order chiral phase transition.

\begin{figure}[thbp]
\epsfxsize=7.5 cm \epsfysize=6.5cm \epsfbox{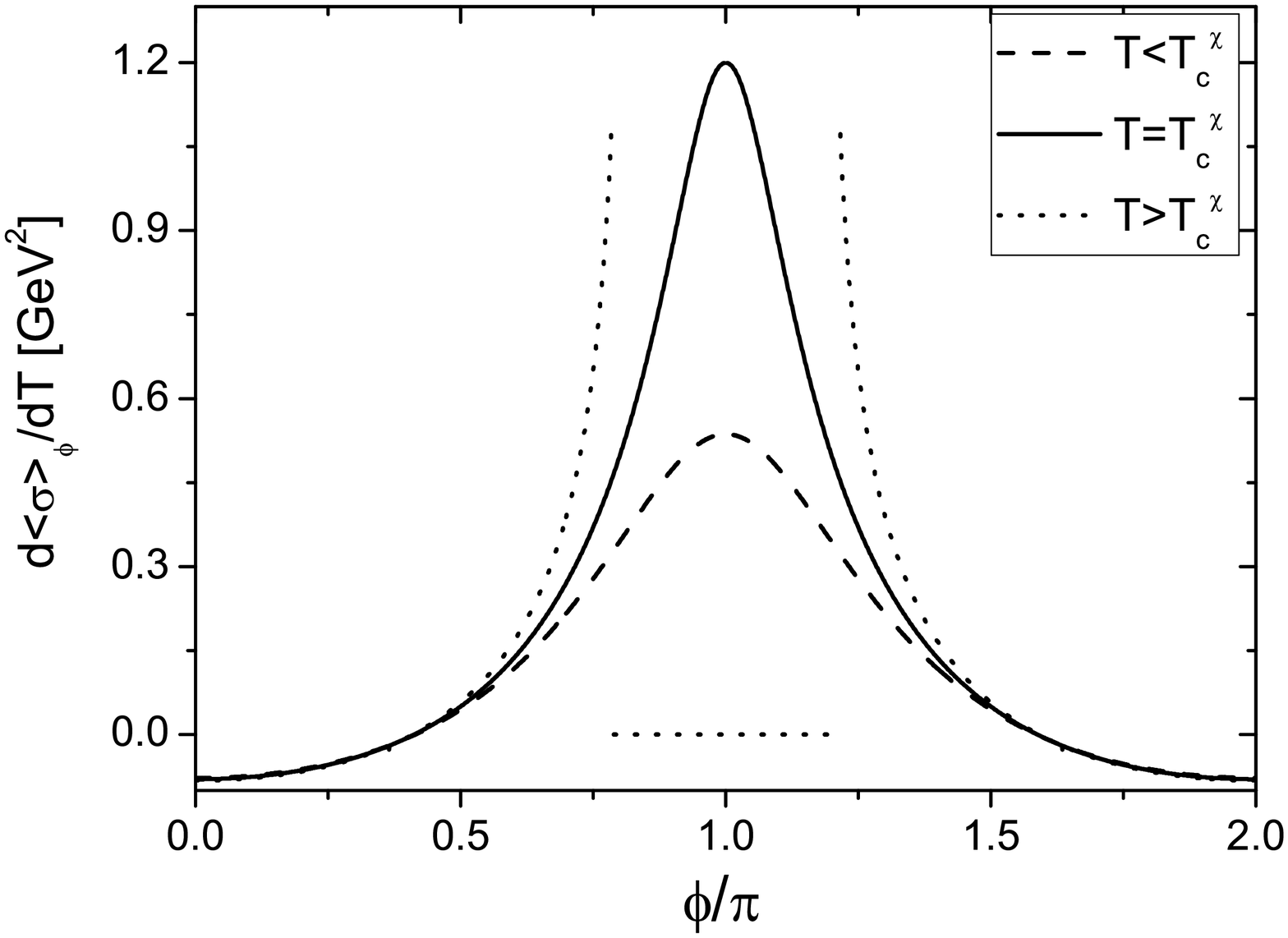}
\caption{Temperature derivative of the general chiral condensate
$\frac{d\langle\sigma\rangle_\phi}{dT}$ for $m=0$ and $\mu=300 {\rm
MeV}$ at three temperature cases: $T < {T_c}^\chi$, $T = {T_c}^\chi$
and $T> {T_c}^\chi$, where ${T_c}^\chi$ is the chiral transition
temperature. }
 \label{fig-firstderi-1order}
\end{figure}
In the case of first order chiral phase transition, the situation is
more complicated. Due to the discontinuity of
$\langle\sigma\rangle_\phi$ at the first order chiral phase
transition point, the temperature derivative of the dressed Polyakov
loop can be expressed as
\begin{equation}
\frac{d\Sigma_1}{dT} =-{\int_0}^{2\pi} \frac{d\phi}{2\pi} e^{-i
\phi} \frac{d\langle \sigma\rangle_\phi}{dT}-\frac{{\rm
Cos}\phi_c}{\pi}\Delta\langle \sigma\rangle_c\frac{d\phi_c}{dT},
\label{eq.dpl1st}
\end{equation}
where, the first term is determined by the regular behavior of
$d\langle \sigma\rangle_\phi/dT$ (see
Fig.\ref{fig-firstderi-1order}), the second term is due to
$\Delta\langle \sigma\rangle_c$, the jump of
$\langle\sigma\rangle_\phi$ at the first order phase transition
point at $\phi=\phi_c$. When $T<T_c^\chi$, the second term vanishes.
Now we consider two limiting cases. First, in the case of a weak
first order phase transition, $\Delta\langle \sigma\rangle_c$ is
small and $d\langle\sigma\rangle_\phi/dT$ around $\phi =\pi$ is
large, as showed in Fig.\ref{fig-firstderi-1order}. So the first
term dominates the result of Eq. (\ref{eq.dpl1st}) and gives the
similar result as that in the case of a second order chiral phase
transition. Second, in the case of a strong first order chiral phase
transition, $\Delta\langle \sigma\rangle_c$ is large and
$d\langle\sigma\rangle_\phi/dT$ is small. So the second term
dominates the result of Eq. (\ref{eq.dpl1st}). Then
$\frac{d\Sigma_1}{dT}$ is strongly dependent on the detailed
information of $d\phi_c/dT$. Our numerical results show that
$d\phi_c/dT$ decreases as temperature increases, and so the second
term also gives a decreasing contribution. In all, it is clear that
$d\Sigma_1/dT$ gets its maximum at $T_c^\chi$, i.e. $T_c^{\cal D}$
and $T_c^\chi$ coincide in the case of a weak first order chiral
phase transition due to remnants of second order chiral phase
transition. The coincidence in the case of a general first order
chiral phase transition is supported by our numerical results and
can be generally expected.

\begin{figure}[thbp]
\epsfxsize=7.5 cm \epsfysize=6.5cm \epsfbox{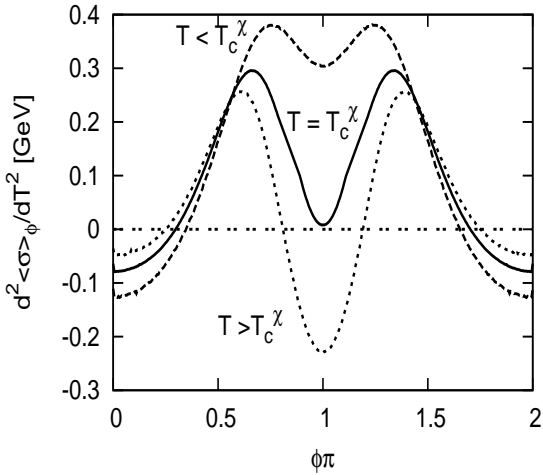}
\caption{Second derivative of the general chiral condensate
$\frac{d^2\langle\sigma\rangle_\phi}{dT^2}$ for $m=5.5{\rm MeV}$ and
$\mu=0$ at three temperature cases: $T < {T_c}^\chi$, $T =
{T_c}^\chi$ and $T> {T_c}^\chi$, where ${T_c}^\chi$ is the chiral
transition temperature. } \label{fig-derivative}
\end{figure}

For the case with finite quark mass and small chemical potential, we
have no phase transitions but crossover. So let us consider the
second temperature derivative of the dressed Polyakov loop
\begin{equation}
\frac{d^2\Sigma_1}{dT^2} =-{\int_0}^{2\pi} \frac{d\phi}{2\pi} e^{-i
\phi} \frac{d^2\langle\sigma\rangle_\phi}{dT^2}, \label{eq.dpl2}
\end{equation}
whose zero point determines the transition temperature $T_c^{\cal
D}$. Fig. \ref{fig-derivative} shows the second derivatives of the
general chiral condensate $d^2\langle\sigma\rangle_\phi/dT^2$ at
three temperature cases: $T < {T_c}^\chi$, $T = {T_c}^\chi$ and $T>
{T_c}^\chi$, where ${T_c}^\chi$ is the chiral transition
temperature. Similar to our previous observation, large values of
$d^2\langle\sigma\rangle_\phi/dT^2$ appear around $\phi=\pi$ (see
the two maximums in  Fig. \ref{fig-derivative}), as remnants of the
second order phase transition in chiral limit. The difference is
that $d^2\langle\sigma\rangle_\phi/dT^2$ in the center region (see
the minimum in  Fig. \ref{fig-derivative}), is suppressed below
${T_c}^\chi$, approaches zero at ${T_c}^\chi$ and changes its sign
above ${T_c}^\chi$. For $T \leq {T_c}^\chi$,
$d^2\langle\sigma\rangle_\phi/dT^2$ around the two maximums dominate
the integral in Eq.(\ref{eq.dpl1}) and $d^2\Sigma_1/dT^2$ does not
change its sign. Above ${T_c}^\chi$, the negative part around the
minimum cancels the contributions from the maximums, and up to a
certain temperature $T_c^{\cal D}$, this cancelation leads to the
zero of the integral in Eq.(\ref{eq.dpl2}). In all, the zero point
of $d^2\Sigma_1/dT^2$ comes from the negative contribution of the
minimum at $T > {T_c}^\chi$, so the transition temperature related
with the dressed Polyakov loop must be higher than the chiral
transition temperature, i.e. ${T_c}^{\cal D}>{T_c}^\chi$.

\section{Phase diagram for three-flavor case}
\label{SU3-section}

We now consider the phase diagram for three-flavor case, and the
parameters are taken from Ref.\cite{Buballa:2003qv,Abuki2010}:

\begin{table}[h]
\begin{tabular}{|c|c|c|c|c|}
\hline & $m_q$[MeV] &\ $m_s$[MeV]  &\ $\;\;G_s\Lambda^2\;\;$ & \
$\;\;K\Lambda^5\;\;$  \\ \hline \ Chiral-limit \ & 0 & 0 & 1.926 &
12.36
\\ \hline \ finite-mass  \ & 5.5  & 140.7  & 1.918 & 12.36    \\
\hline
\end{tabular}
\caption{Two sets of parameters in 3-flavor NJL model: the current
quark mass $m_q$ for up and down quark and $m_s$ for strange quark,
coupling constants $G$ and $K$, with a spatial momentum cutoff
$\Lambda=602.3$ MeV.} \label{tab}
\end{table}

\begin{figure}[thbp]
\epsfxsize=7.5 cm \epsfysize=6.5cm \epsfbox{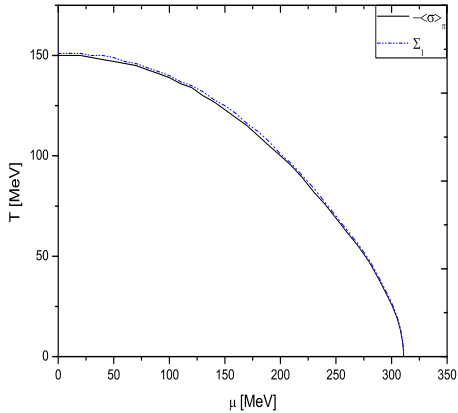}
\caption{Three-flavor phase diagram in the $T-\mu$ plane for the
case of chiral limit. The solid line is the critical line for
$\Sigma_1$, and the dashed line is the critical line for
conventional chiral phase transition. } \label{fig-SU3-chiral-limit}
\end{figure}

In the chiral limit, i.e, for the case of $m_u=m_d=m_s=0$ case, we
find almost exact matching for the transition temperatures
calculated from these two quantities in the whole $T-\mu$ plane as
shown in Fig. \ref{fig-SU3-chiral-limit}. The chiral phase
transition is of 1st order in the whole $T-\mu$ plane.

\begin{figure}[thbp]
\epsfxsize=7.5 cm \epsfysize=6.5cm \epsfbox{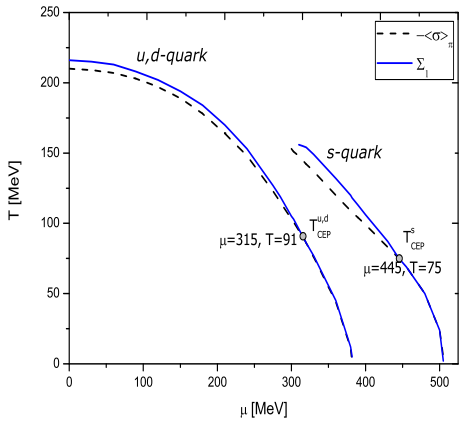}
\caption{Three-flavor phase diagram in the $T-\mu$ plane for the
case of $m_u=m_d=5{\rm MeV}$ and $m_s=140.7{\rm MeV}$. The solid
lines are the critical line for $\Sigma_1$, and the dashed lines are
the critical line for conventional chiral phase transition. The
solid circles indicate the critical end points for chiral phase
transitions of $u,d$ and $s$ quark, respectively.} \label{fig-SU3}
\end{figure}

For the case of finite quark mass $m_u=m_d=5.5 {\rm MeV}$ and
$m_s=140.7{\rm MeV}$, it is observed from Fig. \ref{fig-SU3} that
the the chiral and deconfinement phase transitions are flavor
dependent. For the degenerate light $u,d$ quarks, the result of
phase transitions are similar to that in Fig.
\ref{fig-phasediag-m055}. The critical temperatures of chiral and
deconfinement phase transitions have very small difference in the
low baryon density region when the transition is of crossover, and
the difference vanishes at the critical end point. Here the CEP is
located at $(T_{CEP}^{u,d},\mu_{CEP}^{u,d})=(91{\rm MeV},315{\rm
MeV})$, which is different from Fig. \ref{fig-phasediag-m055}. This
difference comes from: 1) different model parameters have been used,
2) the coupling of $s$ quark to $u,d$ quark contributes one extra
term in the thermodynamical potential comparing with the pure
two-flavor case. For the relative heavy $s$ quark, it is found that
this quark has separate phase transition lines. At low baryon
density, the variation of s quark condensate in the crossover region
is so diffused that it is not possible to identify the crossover
temperature. When density increases, the crossover temperatures for
chiral and deconfinement can be effectively extracted and still
shows the relation of $T_c^{\chi}<T_c^{\cal D}$. These two critical
temperatures of $s$ quark coincide at the critical end point
$(T_{CEP}^{s},\mu_{CEP}^{s})=(75{\rm MeV},445{\rm MeV})$.

\section{Conclusion and discussion}

We investigate the chiral condensate and the dressed Polyakov loop
or dual chiral condensate at finite temperature and density in the
two and three-flavor Nambu--Jona-Lasinio model. We find the behavior
of dressed Polyakov loop in absence of any confinement mechanism
still shows an order parameter like behavior. It is found that in
the chiral limit, the critical temperature for chiral phase
transition coincides with that of the dressed Polyakov loop. In the
case of explicit chiral symmetry breaking, it is found that the
phase transitions are flavor dependent, and the critical temperature
for chiral transition $T_c^{\chi}$ is smaller than that of the
dressed Polyakov loop $T_c^{{\cal D}}$ in the low baryon density
region where the transition is a crossover. With the increase of
current quark mass the difference between the two critical
temperatures is found to be increasing. However, the two critical
temperatures coincide in the high baryon density region where the
phase transition is of first order. For three-flavor case, it is
observed that there are two critical end points in the $(T,\mu)$
phase diagram.

From symmetry analysis, the dressed Polyakov loop can be regarded as
an equivalent order parameter of deconfinement phase transition for
confining theory. In the NJL model, the gluon dynamics is encoded in
a static coupling constant for four point contact interaction. Since
in this work we have included only quark degrees of freedom, a
quantitative comparison will not match with other results. But the
interesting fact is that the qualitative features (angle variation,
temperature variation) of the dressed Polyakov loop remains the
same. Moreover, we expect that independent of the input of
gluedynamics to the quark propagator, it is a general feature for
$T_c^{\chi}=T_c^{\cal D}$ in the case of 1st and 2nd order phase
transitions, and $T_c^{\chi}<T_c^{\cal D}$ in the case of crossover,
which qualitatively agrees with the lattice result in Ref.
\cite{Fodor:2009ax}. This might indicate that for full QCD, in the
crossover case, there exists a small region where chiral symmetry is
restored but the color degrees of freedom are still confined. This
result should be checked in other effective models, e.g. in the
framework of Dyson-Schwinger equations (DSE).

The $(T,\mu)$ phase diagram with three flavors and with $U_A(1)$
anomaly as well as diquark condensate will be studied in the near
future.

\vskip 1cm \noindent {\bf Acknowledgments}: M.H. thanks P. Levai and
local organizers for hospitality, and she also acknowledges
Hungarian Academy of Sciences for supporting the local expenses. The
work of M.H. is supported by CAS program "Outstanding young
scientists abroad brought-in", and NSFC under the number of 10735040
and 10875134, and K.C.Wong Education Foundation, Hong Kong.


\begin{thebibliography}{9}

\bibitem{Polyakov:1978vu}
  A.~M.~Polyakov,
  Phys.\ Lett.\  B {\bf 72}, 477 (1978).

\bibitem{'tHooft:1977hy}
  G.~'t Hooft,
  Nucl.\ Phys.\  B {\bf 138}, 1 (1978).

\bibitem{Casher:1979vw}
  A.~Casher,
  Phys.\ Lett.\  B {\bf 83}, 395 (1979).

\bibitem{Banks:1979yr}
  T.~Banks and A.~Casher,
  Nucl.\ Phys.\  B {\bf 169}, 103 (1980).

\bibitem{Hatta:2003ga}
  Y.~Hatta and K.~Fukushima,
  Phys.\ Rev.\  D {\bf 69}, 097502 (2004).

\bibitem{Mocsy:2003qw}
  A.~Mocsy, F.~Sannino and K.~Tuominen,
  Phys.\ Rev.\ Lett.\  {\bf 92}, 182302 (2004).

\bibitem{Marhauser:2008fz}
  F.~Marhauser and J.~M.~Pawlowski,
  arXiv:0812.1144 [hep-ph].

\bibitem{Braun:2007bx}
  J.~Braun, H.~Gies and J.~M.~Pawlowski,
  Phys.\ Lett.\  B {\bf 684}, 262 (2010).
\bibitem{Braun:2009gm}
  J.~Braun, L.~M.~Haas, F.~Marhauser and J.~M.~Pawlowski,
  arXiv:0908.0008 [hep-ph].

\bibitem{QCD-phase}
T.~Hatsuda and K.~Maeda,
  arXiv:0912.1437 [hep-ph];
K.~Fukushima and T.~Hatsuda,
  Rept.\ Prog.\ Phys.\  {\bf 74}, 014001 (2011);
J.~M.~Pawlowski,
  arXiv:1012.5075 [hep-ph].

\bibitem{McLerran:2007qj}
  L.~McLerran and R.~D.~Pisarski,
  Nucl.\ Phys.\  A {\bf 796}, 83 (2007).

\bibitem{CSC}K.~Rajagopal and F.~Wilczek,
hep-ph/0011333;
D.K.~Hong,
Acta Phys.\ Polon.\ B {\bf 32}, 1253 (2001);
M.~Alford,
Ann.\ Rev.\ Nucl.\ Part.\ Sci.\  {\bf 51}, 131 (2001);
G.~Nardulli,
Riv.\ Nuovo Cim.\  {\bf 25N3}, 1 (2002);
T.~Sch{\"a}fer,
hep-ph/0304281;
H.C.~Ren,
hep-ph/0404074;
M.~Huang,
  Int.\ J.\ Mod.\ Phys.\  E {\bf 14}, 675 (2005);
 I.~A.~Shovkovy,
  Found.\ Phys.\  {\bf 35}, 1309 (2005);
M.~G.~Alford, A.~Schmitt, K.~Rajagopal and T.~Schafer,
  Rev.\ Mod.\ Phys.\  {\bf 80}, 1455 (2008);
Q.~Wang,
arXiv:0912.2485 [nucl-th];

\bibitem{Panero:2009tv}
  M.~Panero,
  Phys.\ Rev.\ Lett.\  {\bf 103}, 232001 (2009).


\bibitem{Kogut:1982rt}
  J.~B.~Kogut, M.~Stone, H.~W.~Wyld, W.~R.~Gibbs, J.~Shigemitsu, S.~H.~Shenker and D.~K.~Sinclair,
  Phys.\ Rev.\ Lett.\  {\bf 50}, 393 (1983).

\bibitem{Fukugita:1986rr}
  M.~Fukugita and A.~Ukawa,
  Phys.\ Rev.\ Lett.\  {\bf 57}, 503 (1986).

\bibitem{Karsch:1994hm}
  F.~Karsch and E.~Laermann,
  Phys.\ Rev.\  D {\bf 50}, 6954 (1994).

\bibitem{Digal:2000ar}
  S.~Digal, E.~Laermann and H.~Satz,
  Eur.\ Phys.\ J.\  C {\bf 18}, 583 (2001).

\bibitem{Digal:2002wn}
  S.~Digal, E.~Laermann and H.~Satz,
  Nucl.\ Phys.\  A {\bf 702}, 159 (2002).


\bibitem{Karsch:2001cy}
  F.~Karsch,
  Lect.\ Notes Phys.\  {\bf 583}, 209 (2002).

\bibitem{Laermann:2003cv}
  E.~Laermann and O.~Philipsen,
  Ann.\ Rev.\ Nucl.\ Part.\ Sci.\  {\bf 53}, 163 (2003).


\bibitem{Bernard:2004je}
  C.~Bernard {\it et al.}  [MILC Collaboration],
  Phys.\ Rev.\  D {\bf 71}, 034504 (2005).

\bibitem{Cheng:2006qk}
  M.~Cheng {\it et al.},
  Phys.\ Rev.\  D {\bf 74}, 054507 (2006).



\bibitem{Aoki:2006br}
  Y.~Aoki, Z.~Fodor, S.~D.~Katz and K.~K.~Szabo,
  Phys.\ Lett.\  B {\bf 643}, 46 (2006).

\bibitem{Aoki:2009sc}
  Y.~Aoki, S.~Borsanyi, S.~Durr, Z.~Fodor, S.~D.~Katz, S.~Krieg and K.~K.~Szabo,
  JHEP {\bf 0906}, 088 (2009).

\bibitem{Fodor:2009ax}
  Z.~Fodor and S.~D.~Katz,
  arXiv:0908.3341 [hep-ph].

\bibitem{Fukushima:2003fw}
  K.~Fukushima,
  Phys.\ Lett.\  B {\bf 591}, 277 (2004).

\bibitem{Ratti:2005jh}
  C.~Ratti, M.~A.~Thaler and W.~Weise,
  Phys.\ Rev.\  D {\bf 73}, 014019 (2006).

\bibitem{Sasaki:2006ww}
 C.~Sasaki, B.~Friman and K.~Redlich,
  Phys.\ Rev.\  D {\bf 75}, 074013 (2007)
  [arXiv:hep-ph/0611147].

\bibitem{Ghosh:2006qh}
  S.~K.~Ghosh, T.~K.~Mukherjee, M.~G.~Mustafa and R.~Ray,
  Phys.\ Rev.\  D {\bf 73}, 114007 (2006).

\bibitem{Fu:2007xc}
W.~j.~Fu, Z.~Zhang and Y.~x.~Liu,
  Phys.\ Rev.\  D {\bf 77}, 014006 (2008).

\bibitem{Zhang:2006gu}
  Z.~Zhang and Y.~X.~Liu,
  Phys.\ Rev.\  C {\bf 75}, 064910 (2007).

\bibitem{Fukushima:2008wg}
  K.~Fukushima,
  Phys.\ Rev.\  D {\bf 77}, 114028 (2008)
  [Erratum-ibid.\  D {\bf 78}, 039902 (2008)].

\bibitem{Abuki:2008nm}
  H.~Abuki, R.~Anglani, R.~Gatto, G.~Nardulli and M.~Ruggieri,
  Phys.\ Rev.\  D {\bf 78}, 034034 (2008).

\bibitem{Schaefer:2007pw}
  B.~J.~Schaefer, J.~M.~Pawlowski and J.~Wambach,
  Phys.\ Rev.\  D {\bf 76}, 074023 (2007).

\bibitem{Mao:2009aq}
  H.~Mao, J.~Jin and M.~Huang,
  J.\ Phys.\ G {\bf 37}, 035001 (2010).

\bibitem{Kondo:2010ts}
  K.~I.~Kondo,
  arXiv:1005.0314 [hep-th].



\bibitem{Synatschke:2007bz}
  F.~Synatschke, A.~Wipf and C.~Wozar,
  Phys.\ Rev.\  D {\bf 75}, 114003 (2007).

\bibitem{Synatschke:2008yt}
  F.~Synatschke, A.~Wipf and K.~Langfeld,
  Phys.\ Rev.\  D {\bf 77}, 114018 (2008).

\bibitem{Bilgici:2008ui}
  E.~Bilgici and C.~Gattringer,
  JHEP {\bf 0805}, 030 (2008).


\bibitem{Gattringer:2006ci}
  C.~Gattringer,
  Phys.\ Rev.\ Lett.\  {\bf 97}, 032003 (2006).

\bibitem{Bruckmann:2006kx}
  F.~Bruckmann, C.~Gattringer and C.~Hagen,
  Phys.\ Lett.\  B {\bf 647}, 56 (2007).

\bibitem{Bilgici:2008qy}
  E.~Bilgici, F.~Bruckmann, C.~Gattringer and C.~Hagen,
  Phys.\ Rev.\  D {\bf 77}, 094007 (2008).


\bibitem{Bruckmann:2008br}
  F.~Bruckmann, C.~Hagen, E.~Bilgici and C.~Gattringer,
  PoS {\bf CONFINEMENT8}, 054 (2008).

\bibitem{Bilgici-thesis} E. Bilgici, PhD Thesis, University of Garz, Austria, 2009
(http://physik.uni-garz.at/itp/files/bilgici/dissertation.pdf).

\bibitem{Fischer:2009wc}
  C.~S.~Fischer,
  Phys.\ Rev.\ Lett.\  {\bf 103}, 052003 (2009).

\bibitem{Fischer:2009gk}
  C.~S.~Fischer and J.~A.~Mueller,
  Phys.\ Rev.\  D {\bf 80}, 074029 (2009).

\bibitem{Fischer:2010fx}
  C.~S.~Fischer, A.~Maas and J.~A.~Mueller,
  arXiv:1003.1960 [hep-ph].

\bibitem{Kashiwa:2009ki}
  K.~Kashiwa, H.~Kouno and M.~Yahiro,
  Phys.\ Rev.\  D {\bf 80}, 117901 (2009).

\bibitem{ourwork} T.~K.~Mukherjee, H.~Chen and M.~Huang,
  Phys.\ Rev.\  D {\bf 82}, 034015 (2010);
F.~K. ~Xu, T. ~K.~Mukherjee, M.~Huang, in preparation.

\bibitem{NJL-report}
U.~Vogl and W.~Weise,
Prog.\ Part.\ Nucl.\ Phys.\  {\bf 27}, 195 (1991);
S.P.~Klevansky,
Rev.\ Mod.\ Phys.\  {\bf 64}, 649 (1992);
T.~Hatsuda and T.~Kunihiro,
Phys.\ Rept.\  {\bf 247}, 221 (1994);
R.~Alkofer, H.~Reinhardt and H.~Weigel,
Phys.\ Rept.\  {\bf 265}, 139 (1996);


\bibitem{Buballa:2003qv}
M.~Buballa,
  Phys.\ Rept.\  {\bf 407}, 205 (2005).

\bibitem{Abuki2010}
H.~Abuki, G.~Baym, T.~Hatsuda and N.~Yamamoto,
 Phys.\ Rev.\  D {\bf 81}, 125010 (2010).
\end{thebibliography}
\end{document}